\newcommand{\g}{$\gamma$}
\newcommand{\cd}{C$_6$D$_6$}

\documentclass[twocolumn]{webofc}
\usepackage{epsfig}
\usepackage{svg}
\usepackage{graphicx}
\DeclareGraphicsExtensions{-eps-converted-to.pdf,.png,.jpg}
\usepackage[varg]{txfonts}   
\usepackage{booktabs}
\usepackage{array} 
\newcolumntype{L}[1]{>{\raggedright\let\newline\\\arraybackslash\hspace{0pt}}m{#1}}
\newcolumntype{C}[1]{>{\centering\let\newline\\\arraybackslash\hspace{0pt}}m{#1}}
\newcolumntype{R}[1]{>{\raggedleft\let\newline\\\arraybackslash\hspace{0pt}}m{#1}}

\graphicspath{{graphics/}{graphics/arch/}{Graphics/}{./}} 
\usepackage{siunitx} 

\begin{document}
\title{Recent highlights and prospects on (n,$\gamma$) measurements at the CERN n\_TOF facility}
%
%

\author{%
J.~Lerendegui-Marco\inst{1} \and %
V.~Alcayne\inst{2} \and %
V.~Babiano-Suarez\inst{1} \and %
M.~Bacak\inst{3} \and %
J.~Balibrea-Correa\inst{1} \and %
A.~Casanovas\inst{4} \and %
C.~Domingo-Pardo\inst{1} \and %
G.~de la Fuente\inst{1} \and %
B.~Gameiro\inst{1}\and %
F.~Garc\'{\i}a-Infantes\inst{5,3}
% ...
\and %
n\_TOF Collaboration}
\institute{%
Instituto de F\'{\i}sica Corpuscular, CSIC - Universidad de Valencia, Spain \and
Centro de Investigaciones Energ\'{e}ticas Medioambientales y Tecnol\'{o}gicas (CIEMAT), Spain \and
European Organization for Nuclear Research (CERN), Switzerland \and
Universitat Polit\`{e}cnica de Catalunya, Spain \and
University of Granada, Spain
% ...
}

\abstract{%
Neutron capture cross-section measurements are fundamental in the study of the slow neutron capture (s-) process of nucleosynthesis and for the development of innovative nuclear technologies. One of the best suited methods to measure radiative neutron capture (n,$\gamma$) cross sections over the full stellar range of interest for all the applications is the time-of-flight (TOF) technique. Overcoming the current experimental limitations for TOF measurements, in particular on low mass unstable samples, requires the combination of facilities with high instantaneous flux, such as the CERN n\_TOF facility, with detection systems with an enhanced detection sensitivity and high counting rate capabilities. This contribution presents a summary about the recent highlights in the field of (n,$\gamma$) measurements at n\_TOF. The recent upgrades in the facility and in new detector concepts for (n,\g) measurements are described. Last, an overview is given on the existing limitations and prospects for TOF measurements involving unstable targets and the outlook for activation measurements at the brand new high-flux n\_TOF-NEAR station.
 }
\maketitle

\section*{Introduction}\label{sec:intro}
Neutron-induced reactions in general, and more particularly radiative neutron capture (n,\g) reactions, are important in various research fields. An improved accuracy of (n,\g) cross sections is required for the design of innovative nuclear devices aimed at the transmutation of nuclear waste, such as accelerator-driven systems, and future Generation IV reactor systems~\cite{Salvatores:08,Colonna:10}.  Neutron capture reactions play also a fundamental role in the slow neutron capture (s-) process of nucleosynthesis operating in red-giant and massive stars~\cite{Kappeler11,Pignatari:10}, which is responsible for the formation of about half of the nuclei heavier than iron. The most crucial and difficult data to obtain are the stellar (n,\g) cross sections of unstable isotopes which act as branchings of the s-process and yield a local isotopic pattern which is very sensitive to the physical conditions of the stellar environment~\cite{Kappeler11}. At present, there are still about 20 relevant s-process branching point isotopes whose cross section could not be measured yet over the neutron energy range of interest for astrophysics due to limitations of the neutron beam facilities, detection systems and attainable sample masses~\cite{Guerrero:17,Domingo:22}.

 Pulsed white neutron beams combined with the time-of-flight (TOF) technique, such as the CERN neutron time-of-flight facility (n\_TOF), are the best suited facilities for comprehensive measurements of neutron capture cross sections. This work reviews some of the recent progresses, existing limitations, and prospects on (n,\g) measurements at the n\_TOF facility. Sec.~\ref{sec:ntof} describes the recent evolution of neutron capture measurements in the n\_TOF facility. The latest upgrades in the facility and in new detection systems are discussed in Sec.~\ref{sec:upgrades}. Last, Sec.~\ref{sec:Perspectives} is focused on the remaining experimental challenges for TOF measurements and the prospects of the new high-flux n\_TOF-NEAR activation station for new measurements involving unstable targets.

\section{Neutron capture measurements at the CERN n\_TOF facility }\label{sec:ntof}

The n\_TOF facility at CERN generates its neutron beams through spallation reactions of 20~GeV/c protons extracted in pulses from the CERN Proton Synchrotron and impinging onto a lead spallation target. The resulting high energy (MeV-GeV) spallation neutrons are partially moderated in a surrounding water layer to produce a white-spectrum neutron beam that expands in energy from thermal to a few GeV. The neutrons travel along different beam lines towards the experimental areas. In neutron capture measurements at n\_TOF, neutrons arriving to the experimental area induce capture reactions in the sample of interest and the emitted $\gamma$-ray cascade is registered by means of dedicated \g-ray detectors.

\begin{figure}[!t]
  \centering
  \includegraphics[width=\columnwidth]{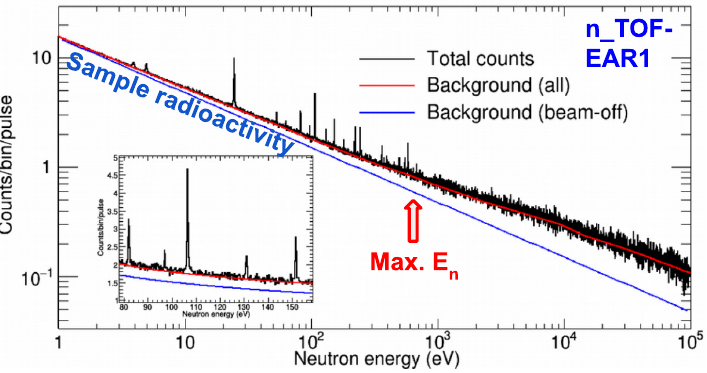}
  \caption{Distribution of counts as a function of the neutron energy
during the $^{171}$Tm(n,$\gamma$) measurement at n\_TOF-EAR1~\cite{Guerrero:2020}. The dominant beam-off background from the activity of the sample limited the observation of  individual resonances to a maximum energy of 700~eV.}
 \label{fig:Tm171}
\end{figure}

Since 2001, more than 60 neutron capture cross section measurements have been carried out at the first experimental area n\_TOF-EAR1~\cite{Guerrero13}. Thanks to its long flight path of 185~m, EAR1 has been a reference facility for high resolution measurements on stable isotopes and samples with sufficient masses ($\geq$100~mg), being especially well suited for the determination of accurate resonance parameters over a broad energy range~\cite{Lerendegui:18,BabianoND:22} and measuring (n,\g) data in the Unresolved Resonance Region (URR) up to 1~MeV~\cite{Aerts:06,Lederer:11,Mingrone:17}. In recent years, measurements on a couple of unstable nuclei $^{63}$Ni,$^{171}$Tm, $^{204}$Tl, acting as s-process branching points~\cite{Kappeler11} were also carried out~\cite{Lederer:13,Guerrero:2020,Casanovas:20}. The low signal-to-background ratio (SBR) due to the neutron-induced background~\cite{Lederer:13} or the activity of the sample~\cite{Guerrero:2020,Casanovas:20}, limited the neutron energy range that could be measured at EAR1. As an example, in the $^{171}$Tm(n,$\gamma$) the dominating background from the activity of the sample allowed to measure resonances only up to 700~eV, as shown in Fig.~\ref{fig:Tm171}.

With the aim of improving the instantaneous luminosity and, as a consequence, the neutron-to-activity ratio for unstable targets, the n\_TOF Collaboration built in 2014 a new vertical beam line, so-called n\_TOF-EAR2~\cite{Weiss15}. This new measuring station, with a flight path of only 20~m, became a world-leading facility in terms of instantaneous flux, about 400 times higher than EAR1~\cite{Lerendegui16}. Over the last years, several experiments have demonstrated its potential for measuring (n,\g) cross sections on highly radioactive and/or small mass samples, for instance $^{244,246,248}$Cm(n,$\gamma$) (600~$\mu$g, 1.8~GBq)~\cite{AlcayneND:22}. On the other hand, the spallation target was not designed to run experiments at EAR2. As a consequence, the energy resolution of this facility was not optimum and was difficult to accurately model it with Monte Carlo (MC) simulations~\cite{AlcayneND:22,Lerendegui:23_ND}. The first (n,\g) measurements at EAR2 also showed the limited performance of the existing \cd~detectors in the high count-rates of EAR2~\cite{AlcayneThesis}. 

\section{Recent upgrades in the facility and detection systems}\label{sec:upgrades}

\begin{figure}[!b]
  \centering
\includegraphics[width=0.96\columnwidth]{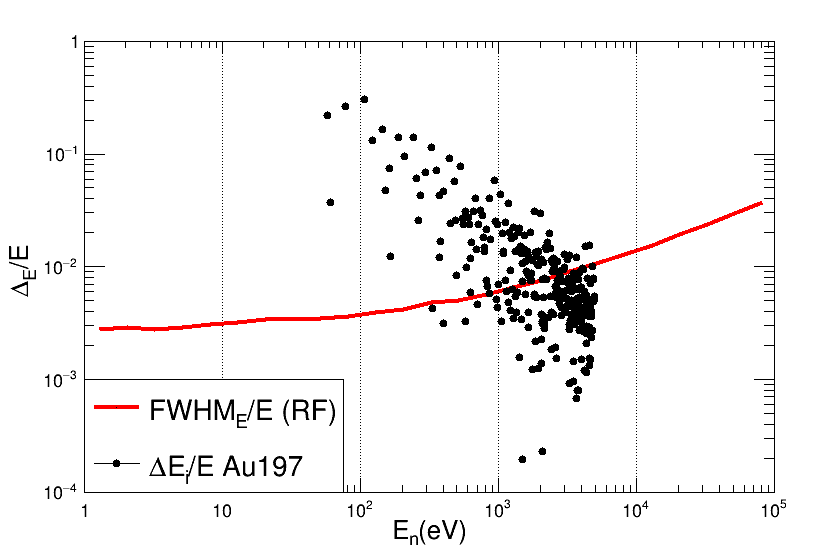}
  \includegraphics[width=0.97\columnwidth]{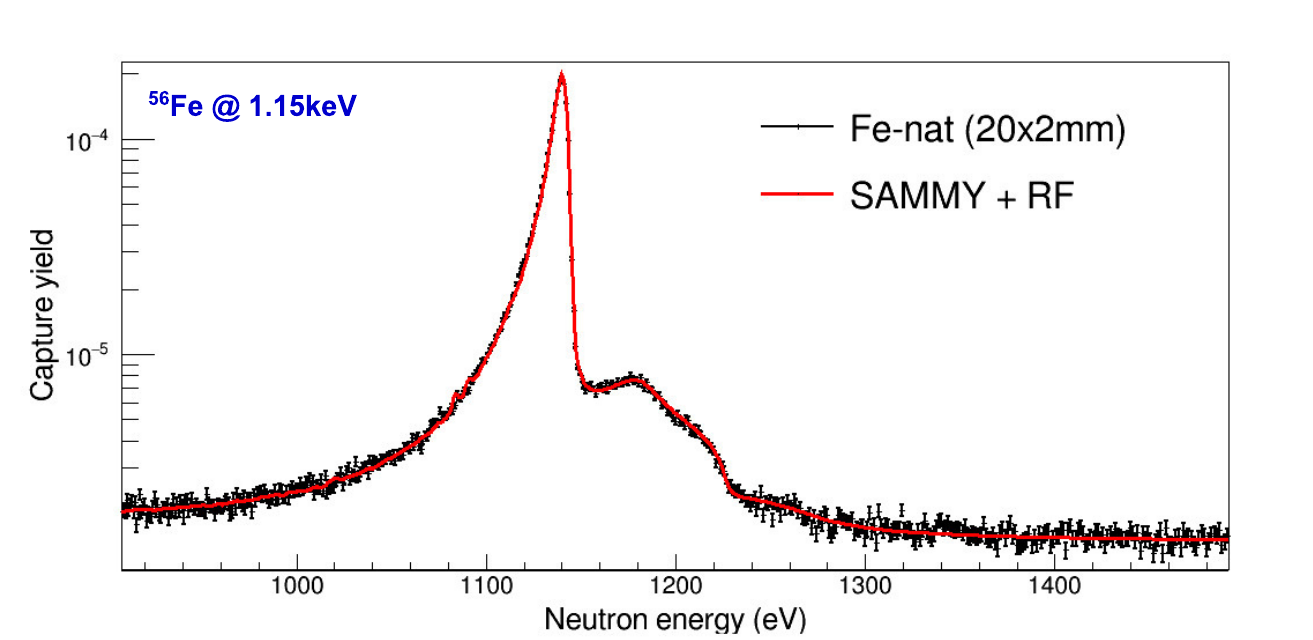}

  \caption{Top: Energy resolution (FWHM) of the upgraded EAR2 facility as a function of the energy compared to the spacings between neighbour resonances of $^{197}$Au(n,\g). Bottom: Resonance of the $^{56}$Fe(n,\g) yield at 1.15~keV measured with at n\_TOF-EAR2 compared to the calculation with SAMMY including the MC-based numerical resolution function.}
 \label{fig:ImprovedRFEAR2}
\end{figure}

The n\_TOF facility installed its third generation spallation target~\cite{Exposito:21} during the CERN Long Shutdown 2 (LS2) (2019-2021) with the aim of optimizing the features of both experimental areas, unlike the previous ones specifically designed for EAR1. The results of the facility commissioning after the upgrade indicate that the excellent energy resolution of EAR1 remains unaltered~\cite{BacakND:22}. As for EAR2, the new spallation target has led to a remarkable improvement in the energy resolution~\cite{Lerendegui:23_ND}. The resulting energy resolution (FWHM) ranges between \SI{3e-3}{}~-~\SI{4e-2}{} in the energy range between 1~eV and 100~keV, which remains, for instance, lower than the resonance spacing for most resonances of $^{197}$Au(n,\g) up to 1~keV, see top panel of Fig.~\ref{fig:ImprovedRFEAR2}). Moreover, after the target upgrade, the resolution function  extracted from MC simulations provides an accurate reproduction of the resonance shapes and energies, as shown in the bottom panel Fig.~\ref{fig:ImprovedRFEAR2}. Lastly, besides the improved resolution, the new target has also enhanced by 30-50\% the neutron flux in both experimental areas~\cite{BacakND:22,PavonND:22}. 

Optimizing the characteristics of the facility is not sufficient to overcome all the existing challenges for (n,\g) measurements. Together with the upgrade of the target, a new generation of detection systems have been also developed, that cope with various challenges that limit the sensitivity of measurements such as the neutron induced background or the high count rates associated to the instantaneous luminosity of EAR2.

 In TOF capture experiments on isotopes with large scattering cross sections relative to capture, a large background component arises from scattered neutrons that get subsequently captured in the surroundings of the \cd~detectors typically used to detect the capture \g-rays, despite their low neutron sensitivity~\cite{Zugec14}. This background has represented the dominant contribution in many (n,\g) experiments in the energy interval of relevance for astrophysics (see e.g. Refs.~\cite{Domingo06,Tagliente13}). In order to suppress this background, a total energy detector with $\gamma$-ray imaging capability, so-called i-TED, has been recently proposed~\cite{Domingo16}. i-TED exploits Compton imaging techniques with the aim of determining the direction of the incoming $\gamma$-rays. This allows rejecting events which do not originate in the sample, thereby enhancing the SBR. This novel detection system has been fully developed and optimized in the recent years~\cite{Babiano20,Balibrea:21}. The imaging-based background rejection capabilities of i-TED were experimentally validated by measuring the $^{56}$Fe(n,$\gamma$) reaction at CERN n\_TOF-EAR1 with an early prototype, achieving an improvement of 3.5 in SBR in the keV range of interest for astrophysics~\cite{Babiano21}.  More details on this device and the recent progresses in its performance can be found in~\cite{Lerendegui:21_IEEE,Lerendegui:22_NIC}. After completing its development, the final i-TED array, consisting of 4 modules, 20 crystals and 1280 readout chanels, has been used in 2022 at n\_TOF EAR1 for the first measurement of the $^{79}$Se($n,\gamma$) capture cross section~\cite{Lerendegui:23_NPA}.

  State-of-the art \cd~detectors (0.6-1 L volume)~\cite{Plag03}, that have been widely and satisfactory used in EAR1, were found to suffer experimental difficulties associated to the high counting rates ($\geq$10~MHz) related to the large luminosity of EAR2. Among others, count-rate and time-of-flight dependent variations of the photo-multiplier (PMT) gain and large pile-up effects have been observed~\cite{AlcayneThesis}. To avoid these issues, the large volume of the existing detectors forces one to place them further away from the sample, thus deteriorating the SBR. To deal with the flux conditions of EAR2, a segmented array of small-volume \cd~detectors, so-called s-TED, has been developed~\cite{AlcayneND:22_b}. Each sTED module contains only 49~ml of scintillation liquid, one order of magnitude less than previous models. Nine of these cells were used in recent capture measurements at EAR2 on the unstable $^{79}$Se(T$_{1/2}$= \SI{3.27e5}{}~y)~\cite{Lerendegui:23_NPA} and $^{94}$Nb(T$_{1/2}$=\SI{2.3e4}{}~y))~\cite{Balibrea:23} in a compact-ring configuration around the capture sample to achieve an optimum efficiency and SBR~\cite{Domingo:22,Lerendegui:23_ND,Lerendegui:23_NPA,Balibrea:23}. This highly segmented and compact setup has triggered further R\&D projects for future detectors based on compact solid scintillators~\cite{Mendoza:23,Balibrea:23b}.


\section{Existing limitations and future perspectives}\label{sec:Perspectives}

\subsection{Prospects on TOF measurements for unstable isotopes of astrophysical interest}\label{sec:TOFunstable}
Neutron capture measurements on unstable isotopes have been already successfully measured at n\_TOF using samples with masses $\simeq$1-50~mg~\cite{Guerrero:2020,Casanovas:20,AlcayneND:22,Lerendegui:23_NPA}. Measurement of lower masses are still difficult, for instance a $^{147}$Pm(n,\g) measurements on a sample of only 85~$\mu$g, just allowed to see the 3 first resonances due to the limited SBR~\cite{Guerrero:19}.

\begin{figure}[!t]
  \centering
  \includegraphics[width=\columnwidth]{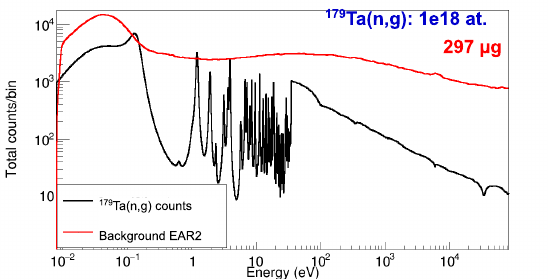}
   \includegraphics[width=0.965\columnwidth]{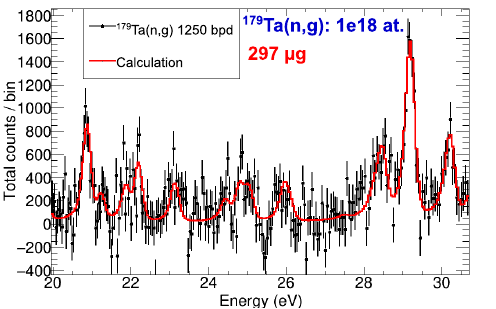} 
  \caption{Top: Expected $^{179}$Ta(n,$\gamma$) counts per pulse as a function of the neutron energy compared to the background level at EAR2, assuming the cross section of TENDL-2021. Bottom: expected capture yield with 1250 bins per decade (bpd) after the background subtraction compared to a calculation, showing that the smaller resonances would be difficult to detect.}
  \label{fig:Ta179_prospects}
\end{figure}

Following the recent aforementioned improvements towards a higher sensitivity in (n,\g) measurements at n\_TOF, we have evaluated the current detection limit of TOF measurements on relevant unstable isotopes in the upgraded n\_TOF-EAR2. Among the challenging key s-process points never measured before via TOF~\cite{Kappeler11,Guerrero:17}, we have selected those that present longer half-lives or lower $\gamma$-ray background from the sample activity -- $^{81}$Kr, $^{135}$Cs, $^{147}$Pm, $^{153}$Gd, $^{163}$Ho, $^{179}$Ta, $^{185}$W. Our feasibility study indicates that with the current signal-to-background ratio of n\_TOF-EAR2, samples with a minimum amount of \SI{5e17}{}~-~\SI{5e18}{} atoms, corresponding to sample masses of only 50-500~$\mu$g, are still required. However, such sample masses are still beyond the maximum attainable quantities for most of the interesting cases. Fig.~\ref{fig:Ta179_prospects} shows the expected results for a potential $^{179}$Ta(n,\g) measurement with a sample of \SI{1e18} atoms. This situation also shows the clear necessity in this field to develop more advanced radiochemical separation techniques and radioanalytical measurement methods, such as those described in Ref.~\cite{Schumann:10}.

\begin{figure}[!b]
  \centering
  \includegraphics[width=\columnwidth]{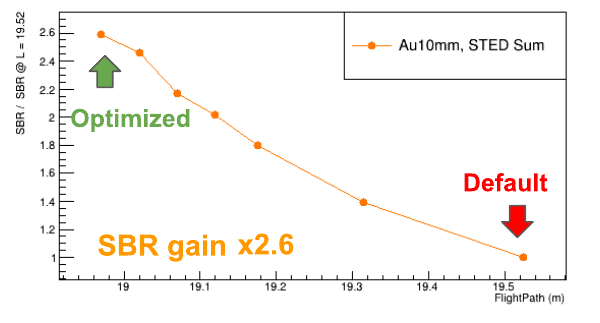}
  \caption{Gain factor in signal-to-background~(1~eV-10~keV) ratio as a function of the flight-path relative to the conditions of the default position (L=19.53~m) used in previous (n,\g) campaigns at EAR2.}
  \label{fig:SBROpt}
\end{figure}

From the viewpoint of the nuclear instrumentation, with the aim of further improving the SBR to access smaller sample mass ($\simeq$10-100~$\mu$g), the new (n,\g) setup at EAR2 has been recently optimized~\cite{INTC_SBR_Opt}. The results of this campaign demonstrate that both the Beam Interception Factor (BIF), defined as the fraction of the neutrons intercepted by the sample, and the SBR can be significantly enhanced by slightly shortening the flight-path (L) of EAR2. In particular, a maximum improvement  in the SBR of factor 2.6 was found for 10~mm diameter samples when lowering the sample and detectors 60~cm with respect to the commonly used setup (see Fig.~\ref{fig:SBROpt}). Furthermore, this campaign also showed that the background is reduced by 20-30\% thanks to the neutron-absorption effect of the lithium polyethylene installed on the floor. This relevant reduction has motivated the design and installation of additional shielding Li-polyethylene layers that is planned in the near future. 

Projects aiming at more striking improvements in the background conditions and the detection sensitivity in TOF measurements for the long-term future are currently being conceptualized. Among others, an enlargement of the EAR or a change of the beam-line optics to avoid the fast neutron component are considered.

\subsection{The n\_TOF-NEAR and CYCLING}\label{sec:NEAR}

\begin{figure}[!h]
  \centering
    \includegraphics[width=0.8\columnwidth]{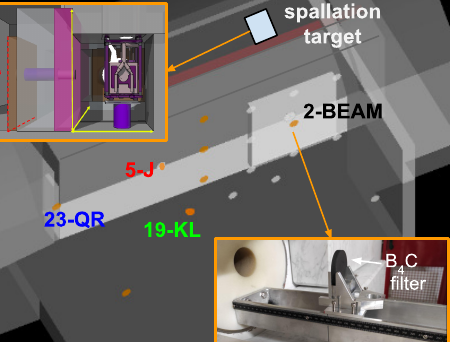}
  \includegraphics[width=\columnwidth]{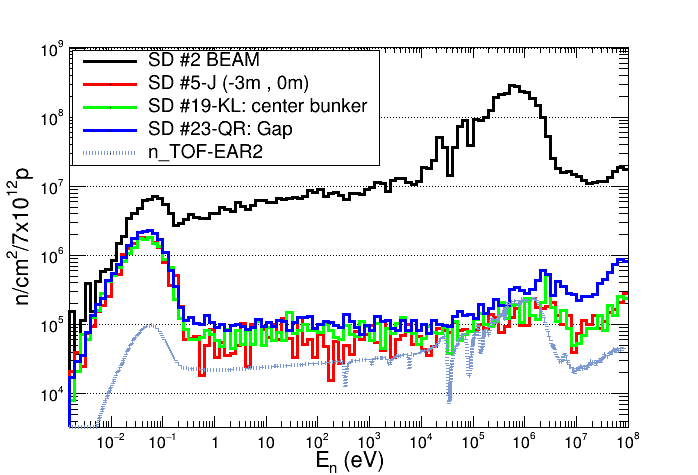}
  \caption{Top: view of the NEAR station as implemented in GEANT4 (top). Bottom:  Neutron fluence of the n\_TOF-EAR2 and NEAR beams compared to the expected off-beam neutron fluence registered in the highlighted sensitive detectors (SD) in the NEAR bunker.}
  \label{fig:NEAR}
\end{figure}

The newest addition to the experimental capabilities of the n\_TOF facility is its recently built NEAR station~\cite{Patronis:23}, at a distance of only 2.5~m from the lead spallation target, shown in the top panel of Fig.~\ref{fig:NEAR}. Its high neutron fluence (see Fig.~\ref{fig:NEAR}), makes it well suited for activation measurements on extremely small mass samples and on radioactive isotopes, in particular (n,$\gamma$) cross-section measurements for nuclear astrophysics which, as discussed in Sec.~\ref{sec:TOFunstable}, are often not feasible via TOF~\cite{Kappeler11}. First calculations indicate that, with a suitable choice of filters (see insert in the top panel of Fig.~\ref{fig:NEAR}), a Maxwellian-like neutron spectrum at stellar temperatures can be generated. This methodology in currently under validation~\cite{StamatiND:22}. 

To date, activation measurements at NEAR consist of week-long irradiations followed by measurements with a high purity Germanium detector (HPGe), far outside NEAR. In this single-activation scheme, the required time to access the area and transport the sample to the offline detector, imposes an intrinsic limit on the minimum half-life of the (n,$\gamma$) product of the accessible physics cases. This limitation can be overcome with the installation of a future cyclic activation station for (n,$\gamma$) experiments (CYCLING)~\cite{loi_background}. This setup would enable the repetition of a short irradiation, rapid transport to a detector, measurement of the decay and transport back to the irradiation position~\cite{Beer_Cyclic}. Depending on the duration of the transport of the sample to the detector, half-lives of the order of seconds become accessible. There are several isotopes of great interest for stellar nucleosynthesis studies that could be investigated with the new CYCLING station~\cite{Domingo:22,Patronis:23}. Some of them are stable nuclei, such as $^{19}$F, which leads to the formation of $^{20}$F with a half-life of only 11~s. Moreover, ($n,\gamma$) measurements on isotopes of relevance for the intermediate ($i$) neutron capture process~\cite{Cowan77} could be accessible for the first time with CYCLING. Some relevant i-process cases could be~\cite{Domingo:22} $^{137}$Cs($n,\gamma$) or $^{144}$Ce($n,\gamma$).

The simplest solution for CYCLING would be the installation of the measuring station inside the NEAR bunker, few meters away from the activation position. However, this poses a substantial challenge due the harsh radiation conditions in the vicinity of the spallation target. In this context, an excellent knowledge of the expected neutron and $\gamma$-ray fields is a prerequisite for its feasibility. MC simulations carried out with FLUKA and Geant4 indicate that the off-beam neutron fluence in the NEAR is similar to that of EAR2's beam (see bottom panel of Fig.~\ref{fig:NEAR}). These results agree within 30\% with the first measurements carried out with CR39 dosimeters in several positions (see top panel of Fig.~\ref{fig:NEAR}). The experimental characterization will be completed with $\gamma$-ray measurements to evaluate the background in-between consecutive bunches. The conclusions of this study will shed light on the feasibility of CYCLING and guide the design of shielding elements that will facilitate the operation of detectors. Alternative options, such as a rabbit system that transports the sample to a decay station outside the NEAR bunker, are also in design phase. 

In the future perspectives of NEAR, new measurements on target-nuclei with short half-lives involved in the $s$- and $i$-processes, could profit from the synergy between NEAR-CYCLING and the nearby ISOLDE facility. A first proposal to produce a  $^{135}$Cs sample of \SI{2.5e15}{} at ISOLDE for a ($n,\gamma$) activation measurement at NEAR has been already accepted~\cite{INTC_Cs135}.

\section{Summary and outlook}\label{sec:summary}
This contribution has presented an overview about the recent upgrades, highlights and future prospects in the field of (n,$\gamma$) measurements at the CERN n\_TOF Facility. 

Since more than 20 years, EAR1 has been an excellent facility for high resolution measurements of stable isotopes and samples with sufficient masses. The opening of the vertical TOF beamline EAR2, with a world-leading instantaneous luminosity, represented a breaktrough for (n,\g) measurements on unstable isotopes with only a few mg samples. More recently, the upgrade of the n\_TOF spallation target has optimized the neutron flux in both experimental areas and has led to a clear improvement of the energy resolution in EAR2, which is key for measuring resolved resonances over a wide range of energies, with better signal-to-background ratio and obtaining accurate resonance parameters.

A large effort has been also dedicated to detector R\&D at n\_TOF to solve existing limitations and further improve the capabilites for (n,\g) measurements. Among them, we should highlight i-TED, an innovative detector concept which applies Compton imaging techniques to suppress the \g-ray large background component induced by scattered neutrons. Moreover, to fully exploit the high instantaneous neutron flux of n\_TOF-EAR2, s-TED, a segmented array of small-volume C$_6$D$_6$ detectors, have been developed. These novel detection systems have been used in the (n,\g) measurements on the unstable isotopes $^{79}$Se and $^{94}$Nb.

Despite recent leading advances, the attainable signal-to-background ratio at n\_TOF-EAR2 is still the limiting factor for the measurement of unstable isotopes with samples below $\simeq$\SI{5e17}{}~atoms. This contribution has discussed the latest optimization campaigns and future ideas aiming at an even enhanced sensitivity. Last, more challenging (n,\g) measurements of short-lived isotopes, for which TOF is not feasible, could be addressed in the new n\_TOF-NEAR facility. Its promising future should be further boosted with the combination of its high flux, the installation of CYCLING and the opportunities offered in the nearby ISOLDE facility to produce suitable unstable samples.

\section*{Acknowledgements}
This work has been supported by the European Research Council (ERC) under the ERC Consolidator Grant project HYMNS, with grant agreement No.~681740 and funding agencies of the participating n\_TOF institutions. Jorge Lerendegui-Marco acknowledges support of grant FJC2020-044688-I funded by MCIN/AEI/ 10.13039/501100011033 and European Union NextGenerationEU/PRTR, and grant CIAPOS/2022/020 funded by the Generalitat Valencia and the European Social Fund. The authors acknowledge support from the Spanish Ministerio de Ciencia e Innovaci\'on under grants PID2019-104714GB-C21 and CSIC under grant CSIC-2023-AEP128.


\end{document}